\documentclass[aps,pra,twocolumn,10pt]{revtex4-1}

\usepackage{amsmath,amssymb,bm}
\usepackage{dsfont}
\usepackage{graphics}

\newcommand{\im}{{i}}
\newcommand{\tr}{\text{tr}}

\newcommand{\Dcirc}{{\cal D}^{\circ}}
\newcommand{\dbar}{d\hspace*{-0.08em}\bar{}\hspace*{0.1em}}

\begin{document}

\title{Stimulated parametric down-conversion for spatiotemporal metrology}
\author{Filippus S. \surname{Roux}}
\email{froux@nmisa.org}
\affiliation{National Metrology Institute of South Africa, Meiring Naud{\'e} Road, Brummeria 0040, Pretoria, South Africa}

\begin{abstract}
A detailed analysis of the stimulated parametric down-conversion (PDC) process is performed to investigate the effects of the spatiotemporal degrees of freedom. The analysis provides information that would be useful for PDC-based metrology applications. Using a Wigner functional approach, we obtain the parametric down-converted state as the Bogoliubov transformed input state, in terms of Bogoliubov kernel functions. The result is used to consider the case for a coherent state seeding stimulated PDC. We also compute the background which is obtained from spontaneous PDC.
\end{abstract}

\maketitle

\section{\label{intro}Introduction}

Parametric down-conversion (PDC) \cite{mandelspdc} has become one of the most widely studied optical phenomena, thanks to its relevance for photonic quantum information systems. Various quantum information applications are based on the PDC process, including the preparation of entangled quantum states \cite{mair,torres1,miatto1,pdcspatcor,entpdc}, preparation of squeezed states \cite{lvovsqu,kolobov,wasil,queez}, and SU(1,1) interferometry \cite{yurke,su11a,su11b}, to name a few.

Here, we focus on metrology applications. Quantum metrology \cite{giovannetti,qmetrev} is usually associated with estimating quantities better than the standard quantum limit \cite{maccone}. However, the association between metrology and quantum physics goes beyond that. While quantum information science can be applied to enhance metrology, standard metrology can also enhance quantum information technology. The latter may play a more significant role in the near future than the former. The reason is that many of the subsystems in quantum information systems require characterization, calibration and standardization to ensure the successful operation of the complete system. Various techniques in quantum optics such as PDC can be used to provide such metrology applications.

One such application is a scheme to measure the absolute intensity of a source. Such a source would be used as the seed optical field in a stimulated PDC experiment, from which the output intensity would be measured. Together with the measured output intensity from the spontaneous PDC in the same experiment --- the output intensity without the seed field --- one should be able to determine the absolute intensity of the seed field. What makes this proposal challenging is the fact that the outputs that are obtained in these two scenarios give very different intensity distributions that depend on the spatiotemporal degrees of freedom in the experiment. Here, we investigate the output intensity distributions that are obtained from stimulated PDC experiments to determine how they depend on the various degrees of freedom.

PDC has been analyzed with diverse formalisms. While many investigations focus on individual photons \cite{mair,arnaut,brambilla0, brambillaspdc,gattispdc,romero,miatto3,perina}, recent advances include bright squeezed states \cite{bsqvac,spdccouple,bright} and non-Gaussian states \cite{kolobov,biswas,lvovsky,trepstheorem}. The use of PDC for metrology applications have been considered in multiphoton cases, with several studies focussing on the temporal frequency aspects of the output state \cite{brecht,triginer}. In contrast, few such studies provide analyses of the full spatiotemporal degrees of freedom of the output state. To address the more challenging analyses of such multiphoton states, a continuous variable formalism is often used \cite{contvar1,weedbrook,contvar2}. It leads to a discretization and truncation of the spatiotemporal degrees of freedom to facilitate the use of Bloch-Messiah reductions \cite{blochmessiah,horoshko}.

Here, we use a Wigner functional approach \cite{entpdc,queez}, which is based on the incorporation of the spatiotemporal degrees of freedom with the particle-number degrees of freedom \cite{stquad,stquaderr,mrowc} to produce a functional phase space for a Moyal representation \cite{groenewold,moyal,psqm} of all quantum optical states. This approach was recently used to derive an evolution equation for the PDC process in a second order nonlinear crystal and to determine the output state obtained from a spontaneous PDC process \cite{nosemi}. In the current investigation, we consider the case where a seed field enters the nonlinear crystal together with the pump, thus producing a stimulated PDC process, often called {\em parametric amplification} or {\em difference frequency generation}, respectively depending on whether the focus is the signal field or the idler field. We'll consider the complete process incorporating both the signal and idler fields. A solution is obtained for an arbitrary quantum state entering as seed field by assuming that the PDC process can be represented as a Bogoliubov transformation of the seed field. We thus find expressions for the Bogoliubov kernels that serve to produce a solution for the output state valid under general conditions.

It is often helpful to use a thin-crystal approximation, which is applicable under typical experimental conditions. We consider two cases: Under some conditions, we can remove all propagation distance dependencies from the kernels, which we call the {\em thin-crystal limit}. In other cases, we retain up to subleading order terms in the propagation distance in the argument of the exponent and the prefactors of the kernels. It is called the {\em thin-crystal approximation}. In the thin crystal limit, we can compute the contributions in the expansions of the kernels to all orders in the squeezing parameter. It shows how successive orders contribute to the output intensity distribution, but some information about the spatial distribution is lost in this limit. The thin-crystal approximation provides a more accurate description of the shape of the output distribution, but, although one can calculate higher order contributions, they become progressively more challenging.

Using the solution for the stimulated PDC process that we obtain as the Bogoliubov transformation of the seed field, we consider the example of a coherent state with a Gaussian parameter function acting as the seed field. This example is studied in the thin crystal limit to see how the different orders contribute to the output intensity distribution. We then use a thin-crystal approximation to determine the conditions for the most efficient difference frequency generation. We also compute the spontaneous PDC background that is produced concurrently with the stimulated PDC output for comparison. With the aid of all these contributions to the output intensity distribution represented in terms of analytic expressions, we then propose a procedure to measure the absolute intensity of the seed field.

\section{\label{bogol}Solving the evolution equation}

Assuming that the pump is a coherent state that remains unentangled with the down-converted light, we can apply the semi-classical approximation. Since the contributions beyond the semi-classical term require un-enhanced vertices, they are severely suppressed relative to the semi-classical term \cite{nosemi}.

Under the semi-classical approximation, the evolution equation for PDC \cite{nosemi} can be written as
\begin{align}
\partial_z W_{\hat{\rho}} = & \frac{1}{2}\alpha^* \diamond H^* \diamond \frac{\delta W_{\hat{\rho}}}{\delta\alpha}
+\frac{1}{2}\frac{\delta W_{\hat{\rho}}}{\delta\alpha^*}\diamond H \diamond \alpha ,
\label{semklas2}
\end{align}
where $W_{\hat{\rho}}$ is the down-converted state's Wigner functional, $H$ is the bilinear semi-classical kernel function for the PDC process, $\alpha$ is the functional phase space field variable, and the $\diamond$-contraction is defined by
\begin{equation}
f\diamond H\diamond g \equiv \int f(\mathbf{k}) H(\mathbf{k},\mathbf{k}')
g(\mathbf{k}')\ \dbar k\ \dbar k' ,
\label{binnespek}
\end{equation}
with
\begin{equation}
\dbar k \equiv \frac{\text{d}^2 k \text{d}\omega}{(2\pi)^3} .
\end{equation}

The semi-classical approximation implies that the phase space field variable for the pump field is replaced by its parameter function. As a consequence, the vertex function for the second-order parametric process is always contracted with this parameter function, so that it becomes the bilinear kernel $H$ in the down-converted degrees of freedom only. It is assumed that the pump parameter function is given by a Gaussian function in the Fourier domain:
\begin{equation}
\zeta(\mathbf{k}) = \sqrt{2\pi} \zeta_0 w_{\text{p}} h(\omega-\omega_{\text{p}},\delta_{\text{p}})
\exp\left(-\tfrac{1}{4} w_{\text{p}}^2 |\mathbf{K}|^2\right) ,
\label{pompprof}
\end{equation}
where $\zeta_0=|\zeta_0|\exp(i\varphi)$ is a complex amplitude, $w_{\text{p}}$ is the beam waist radius, $\mathbf{K}$ is the two-dimensional transverse part of the wave vector $\mathbf{k}$, and $h(\omega)$ is a normalized real-valued spectral function, with a bandwidth $\delta_{\text{p}}$, and a center frequency $\omega_{\text{p}}$. The magnitude of the pump profile function is $\|\zeta(\mathbf{k})\|^2=|\zeta_0|^2$. Under monochromatic conditions, $h^2(\omega-\omega_{\text{p}},\delta_{\text{p}}) \rightarrow 2\pi\delta(\omega-\omega_{\text{p}})$, for $\delta_{\text{p}}\rightarrow 0$.

The expression for the bilinear semi-classical kernel function, obtained by contracting the pump parameter function with the vertex for the PDC, reads \cite{nosemi}
\begin{align}
H(\mathbf{k}_1,\mathbf{k}_2,z)
 = & \frac{-\im 4}{\hbar} \int \zeta^*(\mathbf{k}) T(\mathbf{k}_1,\mathbf{k}_2,\mathbf{k},z)\ \dbar k \nonumber \\
 = & -\im \Omega_0 \sqrt{\omega_1\omega_2} h(\omega_1+\omega_2-\omega_{\text{p}},\delta_{\text{p}}) \nonumber \\
& \times \exp\left(-\tfrac{1}{4} w_{\text{p}}^2 |\mathbf{K}_1+\mathbf{K}_2|^2+\im \Delta k_z z\right) ,
\label{skh}
\end{align}
where $T(\mathbf{k}_1,\mathbf{k}_2,\mathbf{k},z)$ is the vertex function for the second-order parametric process, and
\begin{equation}
\Omega_0 = \frac{4 \sqrt{2\pi\omega_{\text{p}}} \zeta_0^*\sigma_{\text{ooe}}w_{\text{p}}}{c^2} ,
\label{defK0}
\end{equation}
with $\sigma_{\text{ooe}}$ being the nonlinear coefficient of the nonlinear medium for type I phase matching, represented as a scattering cross-section (with the units of an area), and $c$ being the speed of light. The wave vector mismatch $\Delta k_z$ for noncollinear phase matching, is (see Appen.~\ref{dkzaflei})
\begin{align}
\Delta k_z = & \frac{1}{2} \frac{k_z(\omega_1) k_z(\omega_2)}{k_z(\omega_1)+k_z(\omega_2)}
\left|\frac{\mathbf{K}_1}{k_z(\omega_1)}-\frac{\mathbf{K}_2}{k_z(\omega_2)}\right|^2 \nonumber \\
& - \frac{k^2(\omega_1)}{2k_z(\omega_1)} - \frac{k^2(\omega_2)}{2k_z(\omega_2)} + \frac{1}{2}k_z(\omega_1) + \frac{1}{2}k_z(\omega_2) ,
\label{defdkz}
\end{align}
where
\begin{equation}
k_z(\omega) = \frac{\omega \cos[\theta(\omega)]}{v(\omega)} ,  ~~~
k(\omega) = \frac{\omega}{v(\omega)} ,
\label{defkkz}
\end{equation}
with $\theta(\omega)$ being the PDC angle as a function of the frequency and $v(\omega)$ being the weakly dispersive phase velocity in the crystal. The PDC angle is determined by the phase matching condition and is not to be confused with the incident angle of the seed beam. Note that $k_z$ is the $z$-component of the {\em beam axis} and not that of the wave vector of any plane wave. Therefore, it only depends on the frequency and does not depend on the transverse wave vectors.

\subsection{Bogoliubov kernels}

If $W_{\hat{\rho}}[\alpha^*,\alpha](0)$ represents the Wigner functional of the seed field just before it enters the crystal at $z=0$, then the state at $z>0$ is assumed to be represented by a Bogoliubov transformation of the initial seed field. The Bogoliubov transformation changes the arguments of the Wigner functional by
\begin{align}
\begin{split}
\alpha & \rightarrow \bar{\alpha} = U\diamond \alpha + V\diamond \alpha^* \\
\alpha^* & \rightarrow \bar{\alpha}^* = \alpha^*\diamond U^{\dag} + \alpha\diamond V^{\dag} ,
\end{split}
\label{bogoliubov}
\end{align}
where $U$ and $V$ are Bogoliubov kernels such that
\begin{equation}
U^{\dag}\diamond U-V^{\dag}\diamond V=\mathbf{1} .
\end{equation}
Applied to the initial seed field, the Bogoliubov transformation produces
\begin{equation}
W_{\hat{\rho}}[\alpha](z) = W_{\hat{\rho}}[U(z)\diamond\alpha+V(z)\diamond\alpha^*](0) .
\label{bogoseed}
\end{equation}

The Bogoliubov transformed seed field in Eq.~(\ref{bogoseed}) substituted into the evolution equation in Eq.~(\ref{semklas2}) gives
\begin{align}
& \left[\alpha^*\diamond \partial_z U^{\dag}(z)+\alpha\diamond \partial_z V^{\dag}(z)\right] \diamond
\frac{\delta W_{\hat{\rho}}[\bar{\alpha}^*,\bar{\alpha}]}{\delta\bar{\alpha}^*} \nonumber \\
& + \frac{\delta W_{\hat{\rho}}[\bar{\alpha}^*,\bar{\alpha}]}{\delta\bar{\alpha}} \diamond
\left[\partial_z U(z)\diamond\alpha+\partial_z V(z)\diamond\alpha^*\right] \nonumber \\
= & \frac{1}{2}\left[\alpha^* \diamond H^*(z)\diamond V^{\dag}(z)\diamond
\frac{\delta W_{\hat{\rho}}[\bar{\alpha}^*,\bar{\alpha}]}{\delta\bar{\alpha}^*} \right. \nonumber \\
& \left. +\alpha^* \diamond H^*(z)\diamond U^T(z)\diamond
\frac{\delta W_{\hat{\rho}}[\bar{\alpha}^*,\bar{\alpha}]}{\delta\bar{\alpha}} \right. \nonumber \\
& \left. +\frac{\delta W_{\hat{\rho}}[\bar{\alpha}^*,\bar{\alpha}]}{\delta\bar{\alpha}^*}
\diamond U^*(z)\diamond H(z) \diamond \alpha \right. \nonumber \\
& \left. +\frac{\delta W_{\hat{\rho}}[\bar{\alpha}^*,\bar{\alpha}]}{\delta\bar{\alpha}}
\diamond V(z)\diamond H(z) \diamond \alpha \right] .
\end{align}
We can separate the result into four equations, which can then be reduces to two equations:
\begin{align}
\begin{split}
\partial_z U(z) = & \tfrac{1}{2} V(z)\diamond H(z) , \\
\partial_z V(z) = & \tfrac{1}{2} U(z)\diamond H^*(z) .
\end{split}
\label{bogovlgs}
\end{align}

\subsection{Consistency with spontaneous process}

If the Bogoliubov transformed seed state is a solution for the stimulated PDC process, then a Bogoliubov transformed vacuum state should be the solution for the spontaneous PDC process. As a result, the Bogoliubov kernels $U$ and $V$ must combine to produce the squeezed vacuum state kernels $A$ and $B$. It implies that we should be able to use the differential equations for $U$ and $V$, given in Eq.~(\ref{bogovlgs}) to derive differential equations for $A$ and $B$. These latter equations can then be compared with those in \cite{nosemi} to see if they are consistent. It follows that the squeezed vacuum state kernels are represented by the following contractions of the Bogoliubov kernels
\begin{align}
\begin{split}
A = & U^{\dag}\diamond U + V^T\diamond V^* , \\
B = & U^{\dag}\diamond V + V^T\diamond U^* , \\
B^* = & U^T\diamond V^* + V^{\dag}\diamond U .
\end{split}
\label{abuvvac}
\end{align}
We apply a derivative with respect to $z$, substitute Eq.~(\ref{bogovlgs}) into the derivatives of the Bogoliubov kernels, and replace the contractions of Bogoliubov kernels in terms of $A$ and $B$, using Eq.~(\ref{abuvvac}). We also use the fact that $H$ is symmetric. The resulting equations
\begin{align}
\begin{split}
\partial_z A(z) = & \tfrac{1}{2} H^{\dag}(z)\diamond B^*(z) + \tfrac{1}{2} B(z)\diamond H(z) , \\
\partial_z B(z) = & \tfrac{1}{2} H^{\dag}(z)\diamond A^T(z) + \tfrac{1}{2} A(z)\diamond H^*(z) , \\
\partial_z B^*(z) = & \tfrac{1}{2} H^T(z)\diamond  A(z) + \tfrac{1}{2} A^T(z)\diamond H(z) .
\end{split}
\end{align}
are the same as those obtained in \cite{nosemi}, apart from interchanging the definitions of $B$ and $B^*$. It shows that the Bogoliubov solution is consistent with the previous solution in \cite{nosemi}.

\subsection{Solutions for Bogoliubov kernels}

To solve the equations in Eq.~(\ref{bogovlgs}), we can proceed in the same way that was followed in \cite{nosemi}: We integrate the two equations with respect to $z$ and then perform progressive back substitutions to obtain an expansion in terms of integrals of contracted $H$-kernels. The initial conditions for the expansion are assumed to be $U(0)=\mathbf{1}$ and $V(0)=0$, which give the Wigner functional of the initial seed field. The resulting expressions for the Bogoliubov kernels are
\begin{align}
\begin{split}
U(z) = & \mathbf{1}+\frac{1}{4} \int_0^z \int_0^{z_1} H^*(z_2)\diamond H(z_1)\ \text{d}z_2\ \text{d}z_1 \\
& +\frac{1}{16} \int_0^z \int_0^{z_1} \int_0^{z_2} \int_0^{z_3} H^*(z_4)\diamond H(z_3) \\
& \diamond H^*(z_2)\diamond H(z_1)\ \text{d}z_4\ \text{d}z_3\ \text{d}z_2\ \text{d}z_1 ... , \\
V(z) = & \frac{1}{2} \int_0^z H^*(z_1)\ \text{d}z_1
+ \frac{1}{8} \int_0^z \int_0^{z_1} \int_0^{z_2} H^*(z_3) \\
& \diamond H(z_2)\diamond H^*(z_1)\ \text{d}z_3\ \text{d}z_2\ \text{d}z_1 ... .
\end{split}
\label{bogokerne}
\end{align}
Hence, the representation of the stimulated PDC process as a Bogoliubov transformation, produces a successful solution for the evolution equation.

\subsection{Thin-crystal}

Often in experimental implementations of the PDC process, the pump beam has a Rayleigh range that is much longer than the length of the nonlinear crystal. Under such conditions, we can use a {\em thin-crystal} approximation to simplify the expressions.

In the extreme thin-crystal limit, the $z$-dependence is completely removed from the bilinear kernel $H$. The integrals over the $z$'s in the expressions of the Bogoliubov kernels in Eq.~(\ref{bogokerne}) can all be evaluated, leading to
\begin{align}
\begin{split}
U(z) = & \mathbf{1}+\frac{z^2}{2^2 2!} H^*\diamond H +\frac{z^4}{2^4 4!} H^*\diamond H\diamond H^*\diamond H ... \\
= & \mathbf{1}+\sum_{m=1}^{\infty} \frac{z^{2m}}{2^{2m} (2m)!} |H|^{2m\diamond} \\
= & \cosh_{\diamond}\left(\tfrac{1}{2}z|H|\right) , \\
V(z) = & \tfrac{1}{2} H^* + \frac{z^3}{2^3 3!} H^*\diamond H\diamond H^* ... \\
= & \exp(-\im\varphi)\sum_{m=1}^{\infty} \frac{z^{2m-1}}{2^{2m-1} (2m-1)!} |H|^{(2m-1)\diamond} \\
= & \exp(-\im\varphi)\sinh_{\diamond}\left(\tfrac{1}{2}z|H|\right) ,
\end{split}
\label{bogokerne0}
\end{align}
where $H=|H|\exp(\im\varphi)$, with $\varphi$ being the phase of the complex amplitude of the pump parameter function, defined beneath Eq.~(\ref{pompprof}). Moreover, the contractions of the sequences of $H$'s can also be evaluated. The results for odd and even numbers of contracted $H$'s in the thin-crystal limit are
\begin{align}
H_m^{(\text{o})}(\mathbf{k}_1,\mathbf{k}_2)
= & \frac{-\im M_0 M_1^m}{m^{5/4}m!} \exp\left(-\tfrac{1}{4m} w_{\text{p}}^2 |\mathbf{K}_1+\mathbf{K}_2|^2\right)
\omega_1^{\frac{m}{2}} \nonumber \\
& \times \omega_2^{\frac{m}{2}} h(\omega_{\text{p}}-\omega_1-\omega_2,\sqrt{m}\delta_{\text{p}}) ,
\label{onewekern}
\end{align}
where $m$ is an odd integer, and
\begin{align}
H_m^{(\text{e})}(\mathbf{k}_1,\mathbf{k}_2)
 = & \frac{M_0 M_1^m}{m^{5/4}m!}\exp\left(-\tfrac{1}{4m} w_{\text{p}}^2|\mathbf{K}_1-\mathbf{K}_2|^2\right)
 \omega_1^{\frac{m}{2}} \nonumber \\
& \times (\omega_{\text{p}}-\omega_1)^{\frac{m}{2}} h(\omega_1-\omega_2,\sqrt{m}\delta_{\text{p}}) ,
\label{ewekern}
\end{align}
where $m$ is an even integer. The two quantities in the prefactors are given by
\begin{align}
\begin{split}
M_0 = & \frac{\pi^{5/4} w_{\text{p}}^2}{\sqrt{\delta_{\text{p}}}} , \\
M_1 = & \frac{4\sqrt{2}L|\zeta_0|\sigma_{\text{ooe}}\sqrt{\omega_{\text{p}}\delta_{\text{p}}}}{\pi^{3/4} c^2 w_{\text{p}}} ,
\end{split}
\label{defmm}
\end{align}
where $L$ is the length of the nonlinear crystal.

Using the expressions for the odd and even orders in Eqs.~(\ref{onewekern}) and (\ref{ewekern}) with the additional factors of $\tfrac{1}{2}$, we can express the Bogoliubov kernels by
\begin{align}
\begin{split}
U(\mathbf{k}_1,\mathbf{k}_2) = & \mathbf{1}+\sum_{n=1}^{\infty} \frac{1}{4^n}H_{2n}^{(\text{e})}(\mathbf{k}_1,\mathbf{k}_2) , \\
V(\mathbf{k}_1,\mathbf{k}_2) = & \exp(-\im\varphi)\sum_{n=1}^{\infty} \frac{2}{4^n} H_{2n-1}^{(\text{o})}(\mathbf{k}_1,\mathbf{k}_2) .
\end{split}
\label{defuvdkl}
\end{align}

\section{Coherent state seed}

As an illustrative example, we consider the case where the seed field is a coherent state, which is sent into the nonlinear crystal,  together with the coherent state pump.  We apply a Bogoliubov transformation to the arguments of the coherent state of the seed field to obtain the Wigner functional for the stimulated down-converted state exiting the crystal. The result is
\begin{align}
& \mathcal{N}_0 \exp\left( -2 \|\alpha-\xi\|^2 \right) \nonumber \\
 \rightarrow & \mathcal{N}_0 \exp\left(-2\|U\diamond\beta + V\diamond\beta^*-U\diamond\zeta - V\diamond\zeta^*\|^2\right) \nonumber \\
= & \mathcal{N}_0 \exp\left[ -2(\beta^*-\zeta^*)\diamond A\diamond(\beta-\zeta) \right. \nonumber \\
& -(\beta^*-\zeta^*)\diamond B\diamond(\beta^*-\zeta^*) \nonumber \\
& \left. -(\beta-\zeta)\diamond B^*\diamond(\beta-\zeta) \right] ,
\label{kohbogo}
\end{align}
where $\xi$ is the parameter function of the seed field prior to entering the nonlinear crystal, and $\zeta$ is a parameter function whose Bogoliubov transformation produces the original parameter function $\xi$. The combinations of $U$ and $V$ are expressed in terms of the kernels $A$ and $B$ according to Eq.~(\ref{abuvvac}), assuming that $U$ is Hermitian and $V$ is symmetric, and also that $U\diamond V$ is symmetric. These assumed properties of the Bogoliubov kernels allow the definition of an inverse Bogoliubov transformation, which can be used to define $\zeta$ in terms of the original parameter function $\xi$:
\begin{align}
\begin{split}
\zeta & \rightarrow U\diamond \xi - V\diamond \xi^* \\
\zeta^* & \rightarrow \xi^*\diamond U^{\dag} - \xi\diamond V^{\dag} .
\end{split}
\label{zetaxi}
\end{align}
Hence, the effect of a stimulated PDC process on a coherent state is to apply a Bogoliubov transformation on the phase space field variables and an inverse Bogoliubov transformation on the parameter function.

The expression of the Bogoliubov transformed coherent state in Eq.~(\ref{kohbogo}) is that of a displaced squeezed vacuum state. It can now be used to investigate the output intensity distribution that is obtained from stimulated PDC, seeded by a coherent state. For this purpose, we need to consider the measurement process.

\subsection{\label{outint}Output intensity distribution}

To obtain the output intensity distribution that is produced by stimulated PDC with a coherent state as seed field, we perform an intensity measurement on the stimulated PDC state with the aid of a localized number operator that represents the detection process. The Wigner functional for the localized number operator is
\begin{equation}
W_{\hat{n}}[\beta] = \beta^*\diamond D\diamond\beta-\tfrac{1}{2}\tr\{D\} ,
\label{wignum}
\end{equation}
where $D$ is the detector kernel. Placed in the exponent of Eq.~(\ref{kohbogo}), and multiplied by an auxiliary variable $J$, it gives a generating function for the measurement:
\begin{align}
{\cal W}(J) = & \mathcal{N}_0 \int \exp\left[-2\alpha^*\diamond A\diamond\alpha -\alpha^*\diamond B\diamond\alpha^* \right. \nonumber \\
& -\alpha\diamond B^*\diamond\alpha +J(\alpha^*+\zeta^*)\diamond D\diamond(\alpha+\zeta) \nonumber \\
& \left. -\tfrac{1}{2}J\tr\{D\}\right]\ \Dcirc[\alpha] ,
\label{gennum}
\end{align}
where we shifted $\beta\rightarrow \alpha+\zeta$, and $\Dcirc[\alpha]$ is the functional integration measure over the field variable $\alpha$. The generic result of a functional integration of this form is given by
\begin{align}
& \mathcal{N}_0 \int \exp\left(-2\alpha^*\diamond A\diamond\alpha-\alpha^*\diamond B\diamond\alpha^*
-\alpha\diamond B^*\diamond\alpha \right. \nonumber \\
& \left. -\alpha^*\diamond F-F^*\diamond\alpha\right)\ \Dcirc[\alpha] \nonumber \\
  = & \frac{1}{\sqrt{\det\{A\}\det\{A-B\diamond (A^*)^{-1}\diamond B^*\}}} \nonumber \\
& \times \exp\left\{ \tfrac{1}{4} F^*\diamond A^{-1}\diamond F
+ \tfrac{1}{4} \left[F^*-F\diamond (A^*)^{-1}\diamond B^*\right] \right. \nonumber \\
& \diamond\left[A-B\diamond (A^*)^{-1}\diamond B^*\right]^{-1} \nonumber \\
& \left. \diamond \left[F- B\diamond (A^*)^{-1}\diamond F^*\right] \right\} ,
\label{geval13}
\end{align}
where $\mathcal{N}_0$ is a normalization constant. For the expression in Eq.~(\ref{gennum}), we replace $A\rightarrow A-\tfrac{1}{2} J D$, $F^*\rightarrow -J\zeta^*\diamond D$ and $F\rightarrow -JD\diamond\zeta$ in Eq.~(\ref{geval13}), and multiply it by $\exp(J\zeta^*\diamond D\diamond\zeta-\tfrac{1}{2}J\tr\{D\})$. Applying a derivative with respect to $J$ and setting $J=0$, we obtain
\begin{align}
\langle \hat{n} \rangle = & \left. \partial_J {\cal W}(J)\right|_{J=0} \nonumber \\
= & \zeta^*\diamond D\diamond\zeta+\tfrac{1}{2} \tr\{(A-\mathbf{1})\diamond D\} \nonumber \\
& +\tfrac{1}{4} \tr\{A^{-1}\diamond B\diamond A^*\diamond B^*\diamond A^{-1}\diamond D\} \nonumber \\
& -\tfrac{1}{4} \tr\{B\diamond (A^*)^{-1}\diamond B^*\diamond D\} ,
\label{tmpn}
\end{align}
where we used that fact that the argument of a trace can always be transposed and, for a pure squeezed state,
\begin{equation}
A-B\diamond (A^*)^{-1}\diamond B^* = A^{-1} .
\end{equation}
If we can assume that $A$ is real-valued and commutes with $B$, so that
\begin{equation}
A^{-1}\diamond B\diamond A^*\diamond B^*\diamond A^{-1} = A-A^{-1} ,
\label{abkomm}
\end{equation}
then the last two terms in Eq.~(\ref{tmpn}) would cancel. If that is not the case, then at least the leading contributions cancel, leaving terms that are of fourth order in the squeezing parameter \cite{Note1}. Therefore, we consider the expression for the average number of photons in the output to be
\begin{equation}
\langle n \rangle = \zeta^*\diamond D\diamond\zeta+\tfrac{1}{2} \tr\{(A-\mathbf{1})\diamond D\} .
\label{nimage}
\end{equation}
The first term represents the intensity distribution of the stimulated PDC field, where $\zeta$ is given in terms of the parameter function of the seed field by Eq.~(\ref{zetaxi}). The second term is the spontaneous PDC field, which acts as a background noise term.

\subsection{\label{detek}Detector kernel}

\begin{figure}[ht]
\centerline{\includegraphics{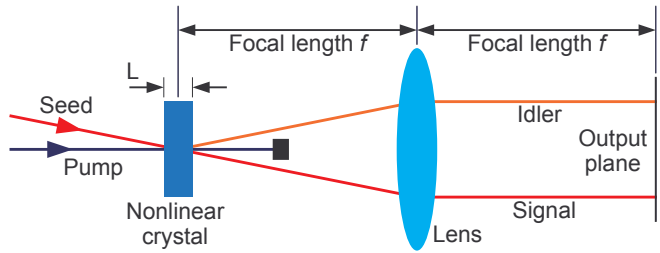}}
\caption{Diagram of the optical setup for the far-field intensity measurement.}
\label{pdcmet}
\end{figure}

To compute the terms in Eq.~(\ref{nimage}), we need to specify the detector kernel, which depends on the details of the experimental setup. For the purpose of a metrology application, we assume that the quantity of interest is obtained from the far-field intensity distribution of the PDC field. It can be obtained with the aid of a $2f$ system, as shown in Fig.~\ref{pdcmet}.

The image of the output intensity distribution is measured with the aid of a CCD array. The detector kernel represents one small detector in the CCD array located at a position $\mathbf{X}_0$ that can vary over the output plane. We'll assume that the detector is smaller than the resolution in the output plane. The $2f$ system performs a Fourier transform of the field distribution in the plane of the nonlinear crystal. The latter field distribution is the inverse Fourier transform of the down-converted state in the Fourier domain. The combination of these two Fourier transformations produces the output intensity distribution of the down-converted state in the Fourier domain via a replacement
\begin{equation}
\mathbf{K} \rightarrow \frac{k_{\text{d}}}{f} \mathbf{X}_0 ,
\label{knax0}
\end{equation}
where $k_{\text{d}}$ is the wavenumber of the down-converted light in the degenerate case, and $f$ is the focal length of the lens in the $2f$ system. The detector includes a narrow spectral filter that fixes the angular frequency to be the degenerate PDC frequency
\begin{equation}
\omega \rightarrow \omega_{\text{d}} = \tfrac{1}{2}\omega_{\text{p}} .
\label{knawd}
\end{equation}

The replacements in Eqs.~(\ref{knax0}) and (\ref{knawd}) can be implemented via Dirac delta functions. However, the resulting kernel may lose idempotency. To retain idempotency, we use a limit process to replace each Dirac delta function with a Gaussian function:
\begin{equation}
\delta(k) = \lim_{w_{\text{D}}\rightarrow\infty} \frac{w_{\text{D}}}{\sqrt{\pi}}\exp(-k^2w_{\text{D}}^2) .
\end{equation}
The result can be seen as a single-mode detector kernel $D(\mathbf{k}_1,\mathbf{k}_2)=M(\mathbf{k}_1)M^*(\mathbf{k}_2)$, where $M(\mathbf{k})$ is a normalized function that is represented by
\begin{align}
M(\mathbf{k}) = & 2\sqrt{2\pi}w_{\text{D}} h(\omega-\omega_{\text{d}},\delta_{\text{D}}) \nonumber \\
& \times \exp\left(-w_{\text{D}}^2 \left|\mathbf{K}-\frac{k_{\text{d}}}{f}\mathbf{X}_0\right|^2\right) .
\end{align}
where $w_{\text{D}}$ is the size of the aperture in the crystal plane that leads to the resolution size in the output plane, and $h(\omega-\omega_{\text{p}},\delta_{\text{D}})$ is the spectral filter function, with a bandwidth $\delta_{\text{D}}$. For a single-mode detector kernel, it follows that $\tr\{D\}=1$. The detector area $\mathcal{A}$ is assumed to be equivalent to the area of the output resolution
\begin{equation}
\mathcal{A} = \frac{2\pi f^2}{k_{\text{d}}^2w_{\text{D}}^2} .
\label{anawd}
\end{equation}

For a large enough $w_{\text{D}}$ and a small enough $\delta_{\text{D}}$, the detector kernel will effectively perform the replacements in Eqs.~(\ref{knax0}) and (\ref{knawd}), so that
\begin{equation}
\zeta^*\diamond D\diamond\zeta = |M^*\diamond\zeta|^2
\rightarrow K_{\text{D}}\left|\zeta\left(\frac{k_{\text{d}}}{f} \mathbf{X}_0,\omega_{\text{d}}\right)\right|^2 ,
\label{uitintens}
\end{equation}
where
\begin{equation}
K_{\text{D}} = \frac{(2\pi)^3\sqrt{\pi}k_{\text{d}}^2\mathcal{A}\delta_{\text{D}}}{f^2} .
\label{kaadee}
\end{equation}
The result in Eq.~(\ref{uitintens}) represents the modulus square of the sum over multiple orders.

\subsection{Transformed parameter function}

Here, we compute the transformed parameter function given in Eq.~(\ref{zetaxi}). The Bogoliubov transformed coherent state is specified in terms of the transformed parameter function and the Bogoliubov kernels.

The inverse Bogoliubov transformed parameter function consists of two terms,
\begin{equation}
\zeta_1 = U\diamond \xi , ~~~
\zeta_2 = V\diamond \xi^* ,
\label{uvxi}
\end{equation}
which we'll compute separately. The complete transformed parameter function is given by $\zeta=\zeta_1-\zeta_2$. The original parameter function is assumed to be a Gaussian spectral function, shifted in the Fourier domain to represent the angle that the seed beam makes with respect to the pump beam as they enter the crystal. It is defined as
\begin{align}
\xi(\mathbf{k}) = & \sqrt{2\pi} \xi_0 w_{\xi} h(\omega-\omega_{\xi},\delta_{\xi})  \nonumber \\
& \times \exp\left(-\tfrac{1}{4} w_{\xi}^2 |\mathbf{K}-\mathbf{K}_{\xi}|^2\right) ,
\label{xiprof}
\end{align}
where $\xi_0$ is the complex amplitude, $w_{\xi}$ is the beam width, $\omega_{\xi}$ is the center frequency, $\delta_{\xi}$ is the bandwidth, and $\mathbf{K}_{\xi}$ is the shift in the spatial Fourier domain. We'll assume that the center frequency is equal to the degenerate frequency $\omega_{\xi}=\omega_{\text{d}} = \tfrac{1}{2}\omega_{\text{p}}$, The associated shift in the output plane is then given by
\begin{equation}
\mathbf{X}_{\xi}=\frac{f}{k_{\text{d}}}\mathbf{K}_{\xi} .
\end{equation}

Next, we calculate the contractions given in Eq.~(\ref{uvxi}). To perform explicit calculations, we'll assume that the experiment satisfies the conditions for the thin-crystal approximation. The calculations are done in two different ways to address different aspects. First, we use the expressions for the Bogoliubov kernels in the thin-crystal limit, as given in Eq.~(\ref{defuvdkl}) in terms of Eqs.~(\ref{onewekern}) and (\ref{ewekern}), to calculated the contractions. It allows us to compute the transformed parameter function to all orders, showing how the different orders contribute to the total intensity distribution and revealing how they change for increasing orders. However, some detailed information about the spatial distribution is lost in the thin-crystal limit. For instance, the background term lacks any spatial degrees of freedom in this limit. For more accurate calculations, we assume that the leading order in the expansions of the kernels in terms of the squeezing parameter dominates. Then we apply a thin crystal approximation that retains some $z$-dependence while allowing a tractable solution. Thus, we obtain a more detailed description of the spatial properties of the output intensity distribution. However, in this case the calculations of higher orders become more challenging. Therefore, we only consider the leading order terms.

\subsection{Thin-crystal limit}

For $\zeta_1$ and $\zeta_2$, respectively, we use the definitions of $U(\mathbf{k}_1,\mathbf{k}_2)$ and $V(\mathbf{k}_1,\mathbf{k}_2)$ in Eq.~(\ref{defuvdkl}) in terms of Eqs.~(\ref{onewekern}) and (\ref{ewekern}), together with Eq.~(\ref{xiprof}). For the spectra $h(\cdot)$, we assume normalized Gaussian functions. The integrals are all readily evaluated. The results are
\begin{align}
\begin{split}
\zeta_1(\mathbf{k}_1) = & \int U(\mathbf{k}_1,\mathbf{k}_2) \xi(\mathbf{k}_2)\ \dbar k_2 \\
= & \xi(\mathbf{k}_1)+\sum_{m=1}^{\infty} \frac{\sqrt{2\pi} \xi_0 w_e^2 \Xi^{2m}
h(\omega-\omega_{\text{d}},\delta_e)}{(2m)!w_{\xi}(1+2m\eta)^{1/4}} \\
& \times \exp\left(-\tfrac{1}{4} w_e^2 |\mathbf{K}_1-\mathbf{K}_{\xi}|^2\right) , \\
\zeta_2(\mathbf{k}_1) = & \int V(\mathbf{k}_1,\mathbf{k}_2) \xi^*(\mathbf{k}_2)\ \dbar k_2 \\
= & \im\sum_{m=1}^{\infty} \frac{\sqrt{2\pi} \xi_0^* w_o^2 \Xi^{2m-1}
h(\omega-\omega_{\text{d}},\delta_o)}{(2m-1)!w_{\xi}(1+2m\eta-\eta)^{1/4}} \\
& \times \exp(-\im\varphi) \exp\left(-\tfrac{1}{4} w_o^2 |\mathbf{K}_1+\mathbf{K}_{\xi}|^2\right) ,
\end{split}
\label{zetaorde}
\end{align}
where
\begin{align}
\begin{split}
w_e = & \frac{w_{\xi} w_{\text{p}}}{\sqrt{w_{\text{p}}^2+2m w_{\xi}^2}} , \\
w_o = & \frac{w_{\xi} w_{\text{p}}}{\sqrt{w_{\text{p}}^2+(2m-1) w_{\xi}^2}} , \\
\delta_e = & \delta_{\xi}\sqrt{1+2m\eta} , \\
\delta_o = & \delta_{\xi}\sqrt{1+2m\eta-\eta} , \\
\eta = & \frac{\delta_{\text{p}}^2}{\delta_{\xi}^2} , \\
\Xi = & \frac{L|\zeta_0|\sigma_{\text{ooe}}\omega_{\text{p}}^{3/2}\sqrt{\delta_{\text{p}}}}{\sqrt{2}\pi^{3/4} c^2 w_{\text{p}}} .
\end{split}
\label{prmdef}
\end{align}
To obtain the output intensity, we perform the replacements in Eqs.~(\ref{knax0}) and (\ref{knawd}), compute the modulus square of $\zeta_1$ and $\zeta_2$ separately, assuming they don't produce overlapping intensity distributions, and multiply the result by $K_{\text{D}}$, as shown in Eq.~(\ref{uitintens}).

\begin{figure}[ht]
\centerline{\includegraphics{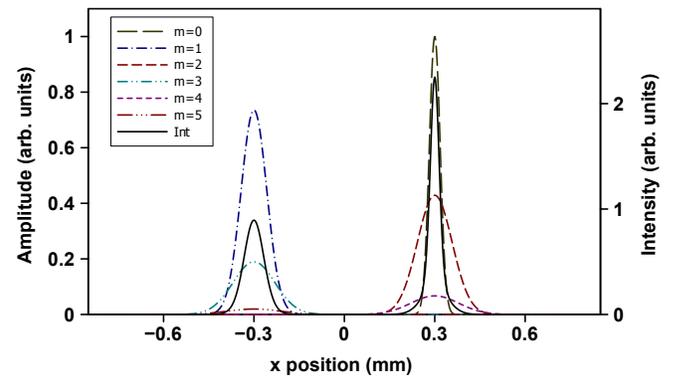}}
\caption{Amplitude distributions for the different orders from $m=0$ to $m=5$ are plotted alons $x$. The intensity of the sum of all the orders is shown by the solid line. The experimental parameters are $\lambda_{\text{d}}=0.8~\mu$m, $f=100$~mm, $w_{\text{p}}=1$~mm, $w_{\xi}=0.6$~mm, $|\mathbf{X}_{\xi}|=0.3$~mm, $\eta=1$, and $\Xi=1.7$. }
\label{dunkris}
\end{figure}

In Fig.~\ref{dunkris}, we plot the one-dimensional curves for the amplitude distributions of the different orders in $\zeta$. We also show the intensity distribution obtained as the modulus square of the sum of all the amplitude distributions. The even orders, which contribute to $\zeta_1$, appear on the right-hand side and the odd orders, which contribute to $\zeta_2$, are on the left-hand side. It shows how the higher order distributions become broader due to the multiple convolutions caused by the multiple contractions of $H$.

Some conclusions can be made from the expressions for the output intensity distribution of the stimulated PDC field and the curves in Fig.~\ref{dunkris}. The signal field $\zeta_1$ that represents the parametric amplified seed field contains the original seed field as a leading contribution. The idler field that represents the difference frequency generated field does not contain any photons from the original seed field. Since it is produced with the complex conjugate of the original parameter function, it represents a natural phase conjugation process. Its leading contribution is a broadened shifted and phase conjugated version of the seed field. The signal and idler fields are shifted in opposite directions due to the opposite signs found in the superposition of the transverse wave vectors in the expressions of the Bogoliubov kernels that are responsible for the signal and idler fields, respectively. The higher order contributions grow progressively broader. The odd numbered higher order terms all contribute to the idler field and the even numbered higher order terms all contribute to the signal field. The expressions for all the orders of the idler field contain global phase factors that are not present in any of the orders of the signal field. Higher orders also come with higher powers of the squeezing parameter $\Xi$. Therefore, in the case of high gain, the contributions of the higher orders will become more prominent, leading to broader intensity distributions.

\subsection{\label{dkb}Thin-crystal approximation}

For the more detailed analysis, we consider only the leading order terms in the signal and idler beams, respectively. The leading order term for the signal beam $\zeta_1$ is the original seed field. The output intensity distribution for this field is directly obtained by applying the replacements to Eq.~(\ref{xiprof}), as in Eq.~(\ref{uitintens}).

The leading order term for the idler beam, generated as the difference frequency beam, is
\begin{align}
\zeta_2(\mathbf{k}_1) = & \frac{1}{2} \int_0^L \int H^*(\mathbf{k}_1,\mathbf{k}_2,z_1) \xi^*(\mathbf{k}_2)\ \dbar k_2\ \text{d}z_1 ,
\end{align}
where $H(\mathbf{k}_1,\mathbf{k}_2,z_1)$ is given in Eq.~(\ref{skh}). The integrations over $\mathbf{k}_2$ can be evaluated readily, but the integration over $z_1$ is challenging. Therefore, we use the thin-crystal approximation to expand the prefactor and the argument of the exponent, respectively, to subleading order in $z_1$. The resulting expression is
\begin{align}
\zeta_2 \approx & \frac{\im \Omega_1 h\left(\omega-\omega_{\text{d}},\sqrt{\delta_{\text{p}}^2+\delta_{\xi}^2}\right)}
{w_0^2 L k_z(\omega_{\text{d}})} \int_0^L \exp\left(-\frac{w_{\text{p}}^2 w_{\xi}^2|\mathbf{K}_a|^2}{4 w_0^2} \right. \nonumber \\
& \left. -\frac{\im z_1 |\mathbf{K}_b|^2}{4k_{\text{d}}w_0^4 }+\im z_1 \chi_{\text{d}} \right)
 \left(k_{\text{d}} w_0^2-\im z_1\right)\ \text{d}z_1 ,
\end{align}
where
\begin{equation}
\Omega_1 = \frac{2\Xi \xi_0^* w_{\xi} w_{\text{p}}^2 \sqrt{2\pi\delta_{\xi}}}
{\left(\delta_{\text{p}}^2+\delta_{\xi}^2\right)^{1/4} w_0^2} ,
\end{equation}
and
\begin{align}
\begin{split}
w_0 = & \sqrt{w_{\text{p}}^2+w_{\xi}^2} , \\
\mathbf{K}_a = & \mathbf{K}_1+\mathbf{K}_{\xi} , \\
\mathbf{K}_b = & \left(2w_{\text{p}}^2+w_{\xi}^2\right)\mathbf{K}_1-w_{\xi}^2\mathbf{K}_{\xi} , \\
\chi_{\text{d}} = & \frac{k^2(\omega_{\text{d}})}{k_z(\omega_{\text{d}})} - k_z(\omega_{\text{d}})
= \frac{k(\omega_{\text{d}}) \sin^2[\theta(\omega_{\text{d}})]}{\cos[\theta(\omega_{\text{d}})]} .
\end{split}
\end{align}

Integrated over $z_1$, the expression reads
\begin{align}
\zeta_2 \approx & \Omega_1 h\left(\omega-\omega_{\text{d}},\sqrt{\delta_{\text{p}}^2+\delta_{\xi}^2}\right)
\exp\left(-\frac{w_{\text{p}}^2 w_{\xi}^2|\mathbf{K}_a|^2}{4 w_0^2}\right)\nonumber \\
& \times \left\{\frac{(\kappa-\beta)}{\kappa^2}[1-\exp(-\im\kappa)]+\frac{\im\beta}{\kappa}\exp(-\im\kappa) \right\} ,
\label{z2dkb}
\end{align}
where
\begin{align}
\begin{split}
\kappa = & \frac{L|\mathbf{K}_b|^2}{4 k_z(\omega_{\text{d}}) w_0^4} - L \chi_{\text{d}} , \\
\beta = & \frac{L}{k_z(\omega_{\text{d}}) w_0^2} .
\end{split}
\end{align}
We recognize the first exponential function in Eq.~(\ref{z2dkb}) as being associated with the original seed field, but shifted in the opposite direction and broadened due to the convolution with the kernel. Since the original field is a Gaussian, the complex conjugation does not have any effect on its shape. This field is modulated by a function of $\kappa$. The effect of this modulation is to vary the peak amplitude of the idler field --- i.e., the efficiency of the difference frequency generation. To see how the efficiency varies, we consider the amplitude at the peak, which is obtained by substituting
\begin{equation}
\mathbf{K}_1\rightarrow -\mathbf{K}_{\xi} ,
\end{equation}
so that $\mathbf{K}_a\rightarrow 0$,
\begin{equation}
\mathbf{K}_b \rightarrow \mathbf{K}_b' = -2 w_0^2 \mathbf{K}_{\xi} ,
\end{equation}
and
\begin{equation}
\kappa \rightarrow a = \frac{L|\mathbf{K}_{\xi}|^2}{k_z(\omega_{\text{d}})} - L \chi_{\text{d}} .
\end{equation}
The dimensionless variable $a$ represents the mismatch in the incident angle, as determined by $\mathbf{K}_{\xi}$, and the PDC angle on which $\chi_{\text{d}}$ depends and which is determined by the phase matching condition. The intensity of the peak is governed by the modulus square of the part of the expression that contains $a$, which is given by the efficiency function
\begin{align}
f(a) = & \frac{\beta^2}{a^2}+\frac{2(a-\beta)\beta\sin a}{a^3} \nonumber \\
& +\frac{2(a-\beta)^2(1-\cos a)}{a^4} .
\label{effunk}
\end{align}
It shows how the efficiency of the difference frequency generation depends on the mismatch between the incident angle and the PDC angle.

\begin{figure}[ht]
\centerline{\includegraphics{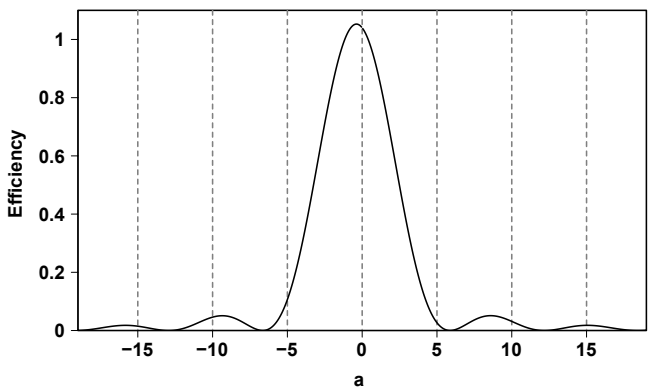}}
\caption{Plot of the efficiency function in Eq.~(\ref{effunk}) for $\beta=0.4$ as a function of the dimensionless variable $a$.}
\label{fa}
\end{figure}

The curve for $f(a)$, shown in Fig.~\ref{fa}, resembles the shape of a sinc-squared function. The location of the peak efficiency, where the most efficient difference frequency generation is obtained, is approximately given by
\begin{equation}
a_{\text{peak}} = -\frac{6\beta}{6+\beta^2} .
\end{equation}
We thus obtain a relationship between the phase matching condition in terms of $\chi(\omega_{\text{d}})$ and the incident angle of the seed beam in terms of $\mathbf{K}_{\xi}$, given by
\begin{align}
\left|\mathbf{K}_{\xi}\right|^2 = & k_z(\omega_{\text{d}}) \chi_{\text{d}} - \frac{k_z(\omega_{\text{d}})}{L} \frac{6\beta}{6+\beta^2} \nonumber \\
\approx & k^2(\omega_{\text{d}}) \sin^2[\theta(\omega_{\text{d}})] - w_0^{-2} .
\label{effek}
\end{align}

In Fig.~\ref{fa}, a relatively large value for $\beta$ is selected to emphasis the shift in the location of the peak efficiency. In most experiments, the value of $\beta$ would be much smaller, moving the peak efficiency closer to the origin.

\subsection{Background term}

The second term in Eq.~(\ref{nimage}) is a background term produced by spontaneous PDC. Usually, the background term would be negligible compared to the term produced by stimulated PDC, due to the enhancement by the number of photons in the seed field. However, we can envisage scenarios where the measurement is made for a small average number of photons in the seed field. Moreover, for absolute intensity measurements, knowledge of this spontaneous PDC term is required. Therefore, we also compute the leading order contribution for the background term.

If an identity operator is used for the detector kernel and the thin-crystal limit is applied to the kernel function $A$, the trace would produce a divergent result. With the localized detector kernel discussed in Section~\ref{detek}, the result is finite, but all the spatial degrees of freedom are lost in the thin-crystal limit, giving a constant background. For a more precise calculation that provides some detail of the spatial behavior of the background, we use the full expression and then apply a thin-crystal approximation to alleviate the calculation.

The leading order term in the expansion of $A-\mathbf{1}$ is second order in the bilinear kernel:
\begin{equation}
A-\mathbf{1} \approx \frac{1}{2} \int_0^L\int_0^L H^*(z_1)\diamond H(z_2)\ \text{d}z_2\ \text{d}z_1 .
\end{equation}
The contraction of the two bilinear kernels produces
\begin{align}
& \int H^*(\mathbf{k}_1,\mathbf{k}_2,z_1) H(\mathbf{k}_2,\mathbf{k}_3,z_2)\ \dbar k_2 \nonumber \\
 = & \Omega_2 h(\omega_1-\omega_3,\sqrt{2} \delta_{\text{p}}) \exp\left[ -R_1 |\mathbf{K}_1-\mathbf{K}_3|^2 \right. \nonumber \\
& \left. - R_2 |\mathbf{K}_1+\mathbf{K}_3|^2 -R_3 \left(|\mathbf{K}_1|^2-|\mathbf{K}_3|^2\right)+R_4\right] ,
\label{hch}
\end{align}
where
\begin{equation}
\Omega_2 = \frac{16 \pi^{5/4} 2^{3/4} \Xi^2 w_{\text{p}}^2 \omega_1 (\omega_{\text{p}}-\omega_1)}
{L^2\omega_{\text{p}}^2\sqrt{\delta_{\text{p}}} \tau(\omega_1,z_1,z_2)} ,
\end{equation}
with $\Xi$ given in Eq.~(\ref{prmdef}), and
\begin{align}
\begin{split}
R_1 = & \frac{w_{\text{p}}^2}{8}
+\frac{\im (z_1-z_2)k_z(\omega_{\text{p}}-\omega_1)}{8k(\omega_{\text{p}})k_z(\omega_1)} \\
 & +\frac{(z_1+z_2)^2}{8 w_{\text{p}}^2k^2(\omega_{\text{p}})\tau(\omega_1,z_1,z_2)} , \\
R_2 = & \frac{\im (z_1-z_2) k(\omega_{\text{p}})}{8 k_z(\omega_1) k_z(\omega_{\text{p}}-\omega_1) \tau(\omega_1,z_1,z_2)} , \\
R_3 = & \frac{\im (z_1+z_2)}{4 k_z(\omega_1)\tau(\omega_1,z_1,z_2)} , \\
R_4 = & \im \tfrac{1}{2}(z_1-z_2) [\chi(\omega_{\text{p}}-\omega_1)+\chi(\omega_1)] ,
\end{split}
\end{align}
with
\begin{align}
\begin{split}
\tau(\omega_1,z_1,z_2) = & 1+\frac{\im (z_1-z_2) k_z(\omega_1)}{w_{\text{p}}^2 k(\omega_{\text{p}}) k_z(\omega_{\text{p}}-\omega_1)} , \\
\chi(\omega) = & \frac{k^2(\omega)}{k_z(\omega)} - k_z(\omega) .
\end{split}
\end{align}

As discussed in Sec.~\ref{detek}, the effect of the detector kernel is to perform the replacements in Eqs.~(\ref{knax0}) and (\ref{knawd}), and multiply the result by $K_{\text{D}}$. Since both $\mathbf{K}_1$ and $\mathbf{K}_3$ are replaced in terms of $\mathbf{X}_0$, the terms with $R_1$ and $R_3$ drop away, leaving only the $R_2$ and $R_4$ terms.

The remaining integrations over $z_1$ and $z_2$ are alleviated by using the thin-crystal approximation as we did in Sec.~\ref{dkb}. Therefore, we use the approximation
\begin{equation}
\frac{1}{\tau(\omega_1,z_1,z_2)} \approx \tau^*(\omega_1,z_1,z_2) .
\end{equation}
The result is
\begin{align}
\begin{split}
\Omega_2 \rightarrow \Omega_2' = & \frac{4 \pi^{5/4} 2^{3/4} \Xi^2 w_{\text{p}}^2}
{L^2 \sqrt{\delta_{\text{p}}}} \tau^*(\omega_1,z_1,z_2) , \\
R_2 \rightarrow R_2' = & \frac{\im (z_1-z_2)}{4 k_z(\omega_{\text{d}})} , \\
R_4 \rightarrow R_4' = & \im (z_1-z_2) \chi_{\text{d}} ,
\end{split}
\end{align}
where we also discarded a $(z_1-z_2)^2$-term in $R_2'$, and used the fact that $k(\omega_{\text{p}})=2k_z(\omega_{\text{d}})$.

\begin{figure}[ht]
\centerline{\includegraphics{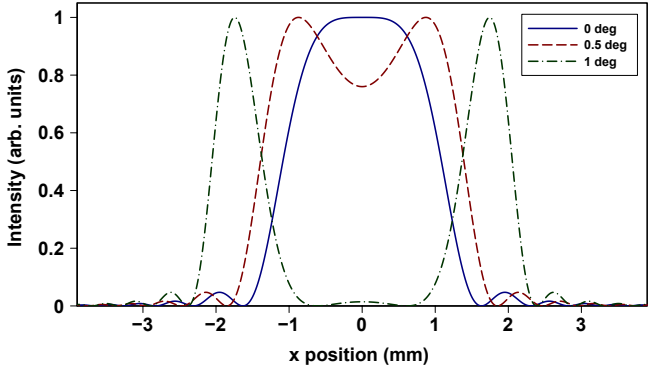}}
\caption{Intensity distribution of the spontaneous PDC background field with $L=3$~mm, $\lambda_{\text{d}}=0.8~\mu$m, $f=100$~mm, and $w_{\text{p}}=0.2$~mm.}
\label{agter}
\end{figure}

After evaluating the $z$-integrations, we obtain
\begin{align}
& \tfrac{1}{2} \tr\{(A-\mathbf{1})\diamond D\} \nonumber \\
= & \Omega_3 \left[\frac{\beta_0 \mathcal{S}}{\left(r^2-r_0^2\right)^2}
+ \frac{2\left(r^2-r_0^2-R^2\right)(1-\mathcal{C})}{\left(r^2-r_0^2\right)^3}\right] ,
\label{uitagt}
\end{align}
where
\begin{align}
\begin{split}
\Omega_3 = & \frac{2 K_{\text{D}} \pi^{3/2} \Xi^2 w_{\text{p}}^2 R^4}{\beta_0^2 \delta_{\text{p}}} , \\
\mathcal{S} = & \sin\left[\frac{\left(r^2-r_0^2\right)\beta_0}{R^2}\right] , \\
\mathcal{C} = & \cos\left[\frac{\left(r^2-r_0^2\right)\beta_0}{R^2}\right] , \\
r = & |\mathbf{X}_0| , \\
r_0 = & f \sin[\theta(\omega_{\text{d}})] , \\
R = & \frac{f}{k_{\text{d}} w_{\text{p}}} , \\
\beta_0 = & \frac{L}{w_{\text{p}}^2 k_{\text{d}} \cos[\theta(\omega_{\text{d}})]} .
\end{split}
\label{defpar}
\end{align}
The output intensity distribution in Eq.~(\ref{uitagt}) is a non-negative real-valued rotationally symmetric function of $r$. The shape of the output intensity distribution is governed by three parameters composed of the experimental parameters.

The intensity distribution of the background field is plotted in Fig.~\ref{agter} along the $x$-axis for three different PDC angles, representing different phase matching conditions. In the experiment, the phase matching condition is changed by rotating the nonlinear crystal to change the angle between the crystal axis and the pump beam axis. For $\theta(\omega_{\text{d}})=0$~degrees, the intensity distribution is a broadened sinc-function. With increasing angle, the peak develops a dip in the center, which eventually produces a ring-shaped intensity distribution.

\subsection{\label{voegsaam}Combined intensity distribution}

\begin{figure}[ht]
\centerline{\includegraphics{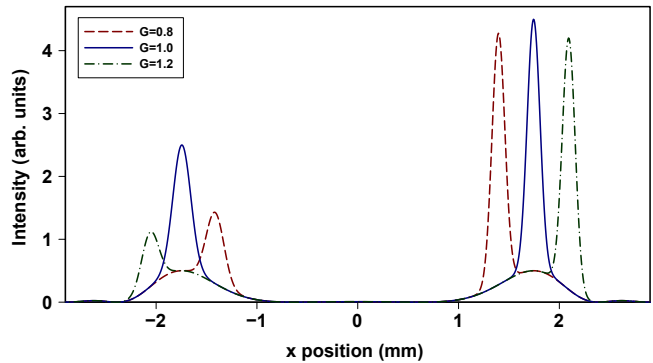}}
\caption{Intensity distribution of the combined output field for three different displacements of the seed field. The experimental parameters: $L=3$~mm, $\lambda_{\text{d}}=0.8~\mu$m, $f=100$~mm, $w_{\text{p}}=w_{\xi}=0.2$~mm, $|\xi_0|^2=4$, and $\Xi=1$.}
\label{saam}
\end{figure}

The combined intensity distribution is obtained by adding the intensities of the signal field, idler field and background field. The result is shown in Fig.~\ref{saam} for a small number of photons in the seed field and for three different angles between the seed and pump beams to demonstrate the effect of the efficiency function. The peaks for the signal field are found on the right-hand side and those for the idler field are on the left-hand side. All the peaks are added on top of the spontaneous PDC background.

The small chosen values for the number of photons in the seed field $|\xi_0|^2$ and the squeezing parameter $\Xi$ allow us to see the spontaneous PDC background together with the stimulated PDC fields. For larger values of these parameters, the stimulated PDC fields would be much larger than the spontaneous PDC background.

Using the quantities defined in Eq.~(\ref{defpar}), we can express the location of the idler peak intensity for the most efficient difference frequency generation in the output plane, based on Eq.~(\ref{effek}), as
\begin{equation}
\left|\mathbf{X}_{\text{peak}}\right|^2 = r_0^2 - \tfrac{1}{2} R^2 .
\end{equation}
A parameter $G$ is used to represent the location of the idler peak intensity in the output plane, according to
\begin{equation}
\mathbf{X}_{\xi} = G \mathbf{X}_{\text{peak}} .
\end{equation}
It can be varied experimentally by changing the incident angle of the seed beam for fixed phase matching conditions. In Fig.~\ref{saam}, we show the curves for $G=0.8, 1, 1.2$. The highest idler peak intensity is obtained for $G=1$, and its location coincides with the location of the peak intensity of the spontaneous PDC background, as expected.

The effective squeezing parameter $\Xi$, given in Eq.~(\ref{prmdef}), serves as the expansion parameter for the different orders of the kernels. In strongly pumped scenarios, the effective squeezing parameter may become quite large. However, the factorials in the denominators in Eq.~(\ref{zetaorde}) ensure that the higher orders will eventually converge. In Fig.~\ref{dunkris}, we used a relatively large value for the effective squeezing parameter so that multiple orders would be visible. For the case in Fig.~\ref{saam} on the other hand, we assumed that the effective squeezing parameter is relatively small so that the leading order would completely dominate.

\section{\label{ansint}Toward absolute intensity}

The fact that the expected output intensity distribution can be reproduced with analytic expressions for the stimulated and spontaneous PDC fields means that the measured output distribution can be used to determine the parameters that govern its shape. By capturing an image of such an output intensity distribution with a CCD array, we can use the analytic expressions to fit the intensity distribution with the unknown parameters as fitting parameters.

Consider for example the case where the intensity of the seed field, which is proportional to the average number of photons $|\xi_0|^2$, is unknown. The three fields shown in Fig.~\ref{saam} --- signal, idler and background --- respectively receive enhancement factors $|\xi_0|^2$, $\Xi|\xi_0|^2$, and $\Xi$, with some minor modifications due to the various bandwidths and beam widths. By fitting the three fields to their respective analytic expressions, the absolute intensity of the seed field can be determined in terms of the average number of photons, even if $\Xi$ is unknown. It may require separate measurements with and without the seed field if the dynamic range of the CCD array cannot see the background together with the stimulated fields.

\section{\label{concl}Conclusions}

The analytic expressions for the output intensity distributions, produced by a stimulated PDC process are obtained with a Wigner functional approach. It produces these results without knowledge of the eigenstates of the process, i.e., without discretizing and truncating the kernel functions of the process. Moreover, it avoids the need for numerical simulations.

This analysis allows the calculation of higher order contributions. However, they do become more challenging as the order increases, unless the thin crystal limit can be applied where all the $z$-dependencies are removed from the expressions of the bilinear kernel. In the thin crystal limit, the information about the phase matching condition is lost. If such information is important, the thin crystal approximation should be used instead, which retains the subleading order $z$-dependencies in the exponents and the prefactors of the kernel functions.

Equipped with these analytic expressions, we may be able to use the measured output intensity distributions from such PDC experiments to determine the values of unknown parameters. It can therefore serve as a tool for metrology applications in radiometry and photometry. One such application is the measurement of the absolute intensity of a light source, which is used as the seed field. It should be possible to use this method for arbitrary large intensities, which would require separate measurements of the stimulated fields and the background field, as well as the calculation of higher order contributions to the output intensity distribution.

\section*{Acknowledgement}

This work was supported in part by funding from the National Research Foundation of South Africa (Grant Numbers: 118532).


\appendix
\section{\label{dkzaflei}Noncollinear phase matching}

Here, we discuss the derivation of the expression for $\Delta k_z$ under paraxial conditions when the PDC angle $\theta(\omega)$ is too large to be considered paraxial. We impose the paraxial condition on each of the beams separately by assuming that the beam divergence angle of each of the three beams is small, even for arbitrary large $\theta(\omega)$.

The pump beam propagates along the $z$-axis, while the signal and idler beams propagate at angles of $\theta(\omega)$ with respect to the $z$-axis. The $z$-components of the wave vectors are given in terms of the angular frequency and the transverse part of the wave vectors by
\begin{equation}
k_z = \sqrt{\frac{\omega^2}{v^2(\omega)} - |\mathbf{K}|^2} .
\end{equation}
To impose the paraxial approximation for the pump beam, it suffices to assume that $|\mathbf{K}|\ll \omega/v(\omega)$. However, since the other two beams propagate at different angles that may be large compared to their beam divergence angles, their situation is more complicated.

For these beams, we replace the transverse part by
\begin{equation}
|\mathbf{K}|^2 \rightarrow \frac{\omega^2\sin^2[\theta(\omega)]}{v^2(\omega)}
+ \left\{|\mathbf{K}|^2 - \frac{\omega^2\sin^2[\theta(\omega)]}{v^2(\omega)}\right\}\epsilon ,
\end{equation}
where we tagged the part in the expression that is small under paraxial conditions by an auxiliary parameter $\epsilon$. We can then substitute the appropriate expressions for the three beams' $z$-components into
\begin{equation}
\Delta k_z = k_z^{(p)}-k_z^{(s)}-k_z^{(i)},
\end{equation}
and expand the result to subleading order in $\epsilon$, after which we set $\epsilon=1$. For critical phase matching, the leading order term cancels, leaving the subleading order which is represented by the expression in Eq.~(\ref{defdkz}).


\begin{thebibliography}{43}
\expandafter\ifx\csname natexlab\endcsname\relax\def\natexlab#1{#1}\fi
\expandafter\ifx\csname bibnamefont\endcsname\relax
  \def\bibnamefont#1{#1}\fi
\expandafter\ifx\csname bibfnamefont\endcsname\relax
  \def\bibfnamefont#1{#1}\fi
\expandafter\ifx\csname citenamefont\endcsname\relax
  \def\citenamefont#1{#1}\fi
\expandafter\ifx\csname url\endcsname\relax
  \def\url#1{\texttt{#1}}\fi
\expandafter\ifx\csname urlprefix\endcsname\relax\def\urlprefix{URL }\fi
\providecommand{\bibinfo}[2]{#2}
\providecommand{\eprint}[2][]{\url{#2}}

\bibitem[{\citenamefont{Hong and Mandel}(1985)}]{mandelspdc}
\bibinfo{author}{\bibfnamefont{C.~K.} \bibnamefont{Hong}} \bibnamefont{and}
  \bibinfo{author}{\bibfnamefont{L.}~\bibnamefont{Mandel}}, ``Theory of
  parametric frequency down conversion of light,'' \bibinfo{journal}{Phys. Rev.
  A} \textbf{\bibinfo{volume}{31}}, \bibinfo{pages}{2409}
  (\bibinfo{year}{1985}).

\bibitem[{\citenamefont{Mair et~al.}(2001)\citenamefont{Mair, Vaziri, Weihs,
  and Zeilinger}}]{mair}
\bibinfo{author}{\bibfnamefont{A.}~\bibnamefont{Mair}},
  \bibinfo{author}{\bibfnamefont{A.}~\bibnamefont{Vaziri}},
  \bibinfo{author}{\bibfnamefont{G.}~\bibnamefont{Weihs}}, \bibnamefont{and}
  \bibinfo{author}{\bibfnamefont{A.}~\bibnamefont{Zeilinger}}, ``Entanglement
  of the orbital angular momentum states of photons,''
  \bibinfo{journal}{Nature} \textbf{\bibinfo{volume}{412}},
  \bibinfo{pages}{313} (\bibinfo{year}{2001}).

\bibitem[{\citenamefont{Torres et~al.}(2003)\citenamefont{Torres, Alexandrescu,
  and Torner}}]{torres1}
\bibinfo{author}{\bibfnamefont{J.~P.} \bibnamefont{Torres}},
  \bibinfo{author}{\bibfnamefont{A.}~\bibnamefont{Alexandrescu}},
  \bibnamefont{and} \bibinfo{author}{\bibfnamefont{L.}~\bibnamefont{Torner}},
  ``Quantum spiral bandwidth of entangled two-photon states,''
  \bibinfo{journal}{Phys. Rev. A} \textbf{\bibinfo{volume}{68}},
  \bibinfo{pages}{050301} (\bibinfo{year}{2003}).

\bibitem[{\citenamefont{Miatto et~al.}(2011)\citenamefont{Miatto, Yao, and
  Barnett}}]{miatto1}
\bibinfo{author}{\bibfnamefont{F.~M.} \bibnamefont{Miatto}},
  \bibinfo{author}{\bibfnamefont{A.~M.} \bibnamefont{Yao}}, \bibnamefont{and}
  \bibinfo{author}{\bibfnamefont{S.~M.} \bibnamefont{Barnett}}, ``{Full
  characterization of the quantum spiral bandwidth of entangled biphotons},''
  \bibinfo{journal}{Phys. Rev. A} \textbf{\bibinfo{volume}{83}},
  \bibinfo{pages}{033816} (\bibinfo{year}{2011}).

\bibitem[{\citenamefont{Walborn et~al.}(2010)\citenamefont{Walborn, Monken,
  P{\'a}dua, and Souto~Ribeiro}}]{pdcspatcor}
\bibinfo{author}{\bibfnamefont{S.~P.} \bibnamefont{Walborn}},
  \bibinfo{author}{\bibfnamefont{C.~H.} \bibnamefont{Monken}},
  \bibinfo{author}{\bibfnamefont{S.}~\bibnamefont{P{\'a}dua}},
  \bibnamefont{and} \bibinfo{author}{\bibfnamefont{P.~H.}
  \bibnamefont{Souto~Ribeiro}}, ``Spatial correlations in parametric
  down-conversion,'' \bibinfo{journal}{Phys. Rep.}
  \textbf{\bibinfo{volume}{495}}, \bibinfo{pages}{87} (\bibinfo{year}{2010}).

\bibitem[{\citenamefont{Roux}(2020{\natexlab{a}})}]{entpdc}
\bibinfo{author}{\bibfnamefont{F.~S.} \bibnamefont{Roux}}, ``Quantifying
  entanglement of parametric down-converted states in all degrees of freedom,''
  \bibinfo{journal}{Phys. Rev. Research} \textbf{\bibinfo{volume}{2}},
  \bibinfo{pages}{023137} (\bibinfo{year}{2020}{\natexlab{a}}).

\bibitem[{\citenamefont{Lvovsky}(2015)}]{lvovsqu}
\bibinfo{author}{\bibfnamefont{A.~I.} \bibnamefont{Lvovsky}}, ``Squeezed
  light,'' \bibinfo{journal}{Photonics: Scientific Foundations, Technology and
  Applications} \textbf{\bibinfo{volume}{1}}, \bibinfo{pages}{121}
  (\bibinfo{year}{2015}).

\bibitem[{\citenamefont{Kolobov}(1999)}]{kolobov}
\bibinfo{author}{\bibfnamefont{M.~I.} \bibnamefont{Kolobov}}, ``The spatial
  behavior of nonclassical light,'' \bibinfo{journal}{Rev. Mod. Phys.}
  \textbf{\bibinfo{volume}{71}}, \bibinfo{pages}{1539} (\bibinfo{year}{1999}).

\bibitem[{\citenamefont{Wasilewski et~al.}(2006)\citenamefont{Wasilewski,
  Lvovsky, Banaszek, and Radzewicz}}]{wasil}
\bibinfo{author}{\bibfnamefont{W.}~\bibnamefont{Wasilewski}},
  \bibinfo{author}{\bibfnamefont{A.~I.} \bibnamefont{Lvovsky}},
  \bibinfo{author}{\bibfnamefont{K.}~\bibnamefont{Banaszek}}, \bibnamefont{and}
  \bibinfo{author}{\bibfnamefont{C.}~\bibnamefont{Radzewicz}}, ``Pulsed
  squeezed light: Simultaneous squeezing of multiple modes,''
  \bibinfo{journal}{Phys. Rev. A} \textbf{\bibinfo{volume}{73}},
  \bibinfo{pages}{063819} (\bibinfo{year}{2006}).

\bibitem[{\citenamefont{Roux}(2021)}]{queez}
\bibinfo{author}{\bibfnamefont{F.~S.} \bibnamefont{Roux}}, ``Spatiotemporal
  effects on squeezing measurements,'' \bibinfo{journal}{Phys. Rev. A}
  \textbf{\bibinfo{volume}{103}}, \bibinfo{pages}{013701}
  (\bibinfo{year}{2021}).

\bibitem[{\citenamefont{Yurke et~al.}(1986)\citenamefont{Yurke, McCall, and
  Klauder}}]{yurke}
\bibinfo{author}{\bibfnamefont{B.}~\bibnamefont{Yurke}},
  \bibinfo{author}{\bibfnamefont{S.~L.} \bibnamefont{McCall}},
  \bibnamefont{and} \bibinfo{author}{\bibfnamefont{J.~R.}
  \bibnamefont{Klauder}}, ``{SU}(2) and {SU}(1,1) interferometers,''
  \bibinfo{journal}{Phys. Rev. A} \textbf{\bibinfo{volume}{33}},
  \bibinfo{pages}{4033} (\bibinfo{year}{1986}).

\bibitem[{\citenamefont{Seyfarth et~al.}(2020)\citenamefont{Seyfarth, Klimov,
  de~Guise, Leuchs, and S{\'a}nchez-Soto}}]{su11a}
\bibinfo{author}{\bibfnamefont{U.}~\bibnamefont{Seyfarth}},
  \bibinfo{author}{\bibfnamefont{A.~B.} \bibnamefont{Klimov}},
  \bibinfo{author}{\bibfnamefont{H.}~\bibnamefont{de~Guise}},
  \bibinfo{author}{\bibfnamefont{G.}~\bibnamefont{Leuchs}}, \bibnamefont{and}
  \bibinfo{author}{\bibfnamefont{L.~L.} \bibnamefont{S{\'a}nchez-Soto}},
  ``Wigner function for {SU}(1,1),'' \bibinfo{journal}{Quantum}
  \textbf{\bibinfo{volume}{4}}, \bibinfo{pages}{317} (\bibinfo{year}{2020}).

\bibitem[{\citenamefont{Ferreri et~al.}(2021)\citenamefont{Ferreri, Santandrea,
  Stefszky, Luo, Herrmann, Silberhorn, and Sharapova}}]{su11b}
\bibinfo{author}{\bibfnamefont{A.}~\bibnamefont{Ferreri}},
  \bibinfo{author}{\bibfnamefont{M.}~\bibnamefont{Santandrea}},
  \bibinfo{author}{\bibfnamefont{M.}~\bibnamefont{Stefszky}},
  \bibinfo{author}{\bibfnamefont{K.~H.} \bibnamefont{Luo}},
  \bibinfo{author}{\bibfnamefont{H.}~\bibnamefont{Herrmann}},
  \bibinfo{author}{\bibfnamefont{C.}~\bibnamefont{Silberhorn}},
  \bibnamefont{and} \bibinfo{author}{\bibfnamefont{P.~R.}
  \bibnamefont{Sharapova}}, ``Spectrally multimode integrated {SU}(1,1)
  interferometer,'' \bibinfo{journal}{Quantum} \textbf{\bibinfo{volume}{5}},
  \bibinfo{pages}{461} (\bibinfo{year}{2021}).

\bibitem[{\citenamefont{Giovannetti et~al.}(2011)\citenamefont{Giovannetti,
  Lloyd, and Maccone}}]{giovannetti}
\bibinfo{author}{\bibfnamefont{V.}~\bibnamefont{Giovannetti}},
  \bibinfo{author}{\bibfnamefont{S.}~\bibnamefont{Lloyd}}, \bibnamefont{and}
  \bibinfo{author}{\bibfnamefont{L.}~\bibnamefont{Maccone}}, ``Advances in
  quantum metrology,'' \bibinfo{journal}{Nature Photon.}
  \textbf{\bibinfo{volume}{5}}, \bibinfo{pages}{222} (\bibinfo{year}{2011}).

\bibitem[{\citenamefont{Polino et~al.}(2020))\citenamefont{Polino, Valeri,
  Spagnolo, and Sciarrino}}]{qmetrev}
\bibinfo{author}{\bibfnamefont{E.}~\bibnamefont{Polino}},
  \bibinfo{author}{\bibfnamefont{M.}~\bibnamefont{Valeri}},
  \bibinfo{author}{\bibfnamefont{N.}~\bibnamefont{Spagnolo}}, \bibnamefont{and}
  \bibinfo{author}{\bibfnamefont{F.}~\bibnamefont{Sciarrino}}, ``Photonic
  quantum metrology,'' \bibinfo{journal}{AVS Quantum Sci.}
  \textbf{\bibinfo{volume}{2}}, \bibinfo{pages}{024703}
  (\bibinfo{year}{2020)}).

\bibitem[{\citenamefont{Giovannetti et~al.}(2004)\citenamefont{Giovannetti,
  Lloyd, and Maccone}}]{maccone}
\bibinfo{author}{\bibfnamefont{V.}~\bibnamefont{Giovannetti}},
  \bibinfo{author}{\bibfnamefont{S.}~\bibnamefont{Lloyd}}, \bibnamefont{and}
  \bibinfo{author}{\bibfnamefont{L.}~\bibnamefont{Maccone}}, ``Quantum-enhanced
  measurements: beating the standard quantum limit,''
  \bibinfo{journal}{Science} \textbf{\bibinfo{volume}{306}},
  \bibinfo{pages}{1330} (\bibinfo{year}{2004}).

\bibitem[{\citenamefont{Arnaut and Barbosa}(2000)}]{arnaut}
\bibinfo{author}{\bibfnamefont{H.~H.} \bibnamefont{Arnaut}} \bibnamefont{and}
  \bibinfo{author}{\bibfnamefont{G.~A.} \bibnamefont{Barbosa}}, ``Orbital and
  intrinsic angular momentum of single photons and entangled pairs of photons
  generated by parametric down-conversion,'' \bibinfo{journal}{Phys. Rev.
  Lett.} \textbf{\bibinfo{volume}{85}}, \bibinfo{pages}{286}
  (\bibinfo{year}{2000}).

\bibitem[{\citenamefont{Brambilla et~al.}(2004)\citenamefont{Brambilla, Gatti,
  Bache, and Lugiato}}]{brambilla0}
\bibinfo{author}{\bibfnamefont{E.}~\bibnamefont{Brambilla}},
  \bibinfo{author}{\bibfnamefont{A.}~\bibnamefont{Gatti}},
  \bibinfo{author}{\bibfnamefont{M.}~\bibnamefont{Bache}}, \bibnamefont{and}
  \bibinfo{author}{\bibfnamefont{L.~A.} \bibnamefont{Lugiato}}, ``Simultaneous
  near-field and far-field spatial quantum correlations in the high-gain regime
  of parametric down-conversion,'' \bibinfo{journal}{Phys. Rev. A}
  \textbf{\bibinfo{volume}{69}}, \bibinfo{pages}{023802}
  (\bibinfo{year}{2004}).

\bibitem[{\citenamefont{Brambilla et~al.}(2010)\citenamefont{Brambilla,
  Caspani, Lugiato, and Gatti}}]{brambillaspdc}
\bibinfo{author}{\bibfnamefont{E.}~\bibnamefont{Brambilla}},
  \bibinfo{author}{\bibfnamefont{L.}~\bibnamefont{Caspani}},
  \bibinfo{author}{\bibfnamefont{L.~A.} \bibnamefont{Lugiato}},
  \bibnamefont{and} \bibinfo{author}{\bibfnamefont{A.}~\bibnamefont{Gatti}},
  ``Spatiotemporal structure of biphoton entanglement in type-II parametric
  down-conversion,'' \bibinfo{journal}{Phys. Rev. A}
  \textbf{\bibinfo{volume}{82}}, \bibinfo{pages}{013835}
  (\bibinfo{year}{2010}).

\bibitem[{\citenamefont{Gatti et~al.}(2012)\citenamefont{Gatti, Corti,
  Brambilla, and Horoshko}}]{gattispdc}
\bibinfo{author}{\bibfnamefont{A.}~\bibnamefont{Gatti}},
  \bibinfo{author}{\bibfnamefont{T.}~\bibnamefont{Corti}},
  \bibinfo{author}{\bibfnamefont{E.}~\bibnamefont{Brambilla}},
  \bibnamefont{and} \bibinfo{author}{\bibfnamefont{D.~B.}
  \bibnamefont{Horoshko}}, ``Dimensionality of the spatiotemporal entanglement
  of parametric down-conversion photon pairs,'' \bibinfo{journal}{Phys. Rev. A}
  \textbf{\bibinfo{volume}{86}}, \bibinfo{pages}{053803}
  (\bibinfo{year}{2012}).

\bibitem[{\citenamefont{Romero et~al.}(2012)\citenamefont{Romero, Giovannini,
  Franke-Arnold, Barnett, and Padgett}}]{romero}
\bibinfo{author}{\bibfnamefont{J.}~\bibnamefont{Romero}},
  \bibinfo{author}{\bibfnamefont{D.}~\bibnamefont{Giovannini}},
  \bibinfo{author}{\bibfnamefont{S.}~\bibnamefont{Franke-Arnold}},
  \bibinfo{author}{\bibfnamefont{S.~M.} \bibnamefont{Barnett}},
  \bibnamefont{and} \bibinfo{author}{\bibfnamefont{M.~J.}
  \bibnamefont{Padgett}}, ``Increasing the dimension in high-dimensional
  two-photon orbital angular momentum entanglement,'' \bibinfo{journal}{Phys.
  Rev. A} \textbf{\bibinfo{volume}{86}}, \bibinfo{pages}{012334}
  (\bibinfo{year}{2012}).

\bibitem[{\citenamefont{Miatto et~al.}(2012)\citenamefont{Miatto, Pires,
  Barnett, and van Exter}}]{miatto3}
\bibinfo{author}{\bibfnamefont{F.~M.} \bibnamefont{Miatto}},
  \bibinfo{author}{\bibfnamefont{H.~D.~L.} \bibnamefont{Pires}},
  \bibinfo{author}{\bibfnamefont{S.~M.} \bibnamefont{Barnett}},
  \bibnamefont{and} \bibinfo{author}{\bibfnamefont{M.~P.} \bibnamefont{van
  Exter}}, ``Spatial {S}chmidt modes generated in parametric down-conversion,''
  \bibinfo{journal}{Eur. Phys. J. D} \textbf{\bibinfo{volume}{66}},
  \bibinfo{pages}{263} (\bibinfo{year}{2012}).

\bibitem[{\citenamefont{Perina~Jr.}(2019)}]{perina}
\bibinfo{author}{\bibfnamefont{J.}~\bibnamefont{Perina~Jr.}}, ``Waves in
  spatio-spectral and temporal coherence of evolving ultra-intense twin
  beams,'' \bibinfo{journal}{Sci. Rep.} \textbf{\bibinfo{volume}{9}},
  \bibinfo{pages}{4256} (\bibinfo{year}{2019}).

\bibitem[{\citenamefont{Chekhova et~al.}(2015)\citenamefont{Chekhova, Leuchs,
  and {\.Z}ukowski}}]{bsqvac}
\bibinfo{author}{\bibfnamefont{M.~V.} \bibnamefont{Chekhova}},
  \bibinfo{author}{\bibfnamefont{G.}~\bibnamefont{Leuchs}}, \bibnamefont{and}
  \bibinfo{author}{\bibfnamefont{M.}~\bibnamefont{{\.Z}ukowski}}, ``Bright
  squeezed vacuum: Entanglement of macroscopic light beams,''
  \bibinfo{journal}{Opt. Commun.} \textbf{\bibinfo{volume}{337}},
  \bibinfo{pages}{27} (\bibinfo{year}{2015}).

\bibitem[{\citenamefont{Barral et~al.}(2017)\citenamefont{Barral, Belabas,
  Procopio, d'Auria, Tanzilli, Bencheikh, and Levenson}}]{spdccouple}
\bibinfo{author}{\bibfnamefont{D.}~\bibnamefont{Barral}},
  \bibinfo{author}{\bibfnamefont{N.}~\bibnamefont{Belabas}},
  \bibinfo{author}{\bibfnamefont{L.~M.} \bibnamefont{Procopio}},
  \bibinfo{author}{\bibfnamefont{V.}~\bibnamefont{d'Auria}},
  \bibinfo{author}{\bibfnamefont{S.}~\bibnamefont{Tanzilli}},
  \bibinfo{author}{\bibfnamefont{K.}~\bibnamefont{Bencheikh}},
  \bibnamefont{and} \bibinfo{author}{\bibfnamefont{J.~A.}
  \bibnamefont{Levenson}}, ``Continuous-variable entanglement of two bright
  coherent states that never interacted,'' \bibinfo{journal}{Phys. Rev. A}
  \textbf{\bibinfo{volume}{96}}, \bibinfo{pages}{053822}
  (\bibinfo{year}{2017}).

\bibitem[{\citenamefont{Sharapova et~al.}(2020)\citenamefont{Sharapova,
  Frascella, Riabinin, P{\'e}rez, Tikhonova, Lemieux, Boyd, Leuchs, and
  Chekhova}}]{bright}
\bibinfo{author}{\bibfnamefont{P.~R.} \bibnamefont{Sharapova}},
  \bibinfo{author}{\bibfnamefont{G.}~\bibnamefont{Frascella}},
  \bibinfo{author}{\bibfnamefont{M.}~\bibnamefont{Riabinin}},
  \bibinfo{author}{\bibfnamefont{A.~M.} \bibnamefont{P{\'e}rez}},
  \bibinfo{author}{\bibfnamefont{O.~V.} \bibnamefont{Tikhonova}},
  \bibinfo{author}{\bibfnamefont{S.}~\bibnamefont{Lemieux}},
  \bibinfo{author}{\bibfnamefont{R.~W.} \bibnamefont{Boyd}},
  \bibinfo{author}{\bibfnamefont{G.}~\bibnamefont{Leuchs}}, \bibnamefont{and}
  \bibinfo{author}{\bibfnamefont{M.~V.} \bibnamefont{Chekhova}}, ``Properties
  of bright squeezed vacuum at increasing brightness,'' \bibinfo{journal}{Phys.
  Rev. Research} \textbf{\bibinfo{volume}{2}}, \bibinfo{pages}{013371}
  (\bibinfo{year}{2020}).

\bibitem[{\citenamefont{Biswas and Agarwal}(2007)}]{biswas}
\bibinfo{author}{\bibfnamefont{A.}~\bibnamefont{Biswas}} \bibnamefont{and}
  \bibinfo{author}{\bibfnamefont{G.~S.} \bibnamefont{Agarwal}},
  ``Nonclassicality and decoherence of photon-subtracted squeezed states,''
  \bibinfo{journal}{Phys. Rev. A} \textbf{\bibinfo{volume}{75}},
  \bibinfo{pages}{032104} (\bibinfo{year}{2007}).

\bibitem[{\citenamefont{Lvovsky et~al.}(2020)\citenamefont{Lvovsky, Grangier,
  Ourjoumtsev, Parigi, Sasaki, and Tualle-Brouri}}]{lvovsky}
\bibinfo{author}{\bibfnamefont{A.~I.} \bibnamefont{Lvovsky}},
  \bibinfo{author}{\bibfnamefont{P.}~\bibnamefont{Grangier}},
  \bibinfo{author}{\bibfnamefont{A.}~\bibnamefont{Ourjoumtsev}},
  \bibinfo{author}{\bibfnamefont{V.}~\bibnamefont{Parigi}},
  \bibinfo{author}{\bibfnamefont{M.}~\bibnamefont{Sasaki}}, \bibnamefont{and}
  \bibinfo{author}{\bibfnamefont{R.}~\bibnamefont{Tualle-Brouri}}, ``Production
  and applications of non-gaussian quantum states of light,''
  \bibinfo{journal}{arXiv preprint arXiv:2006.16985}  (\bibinfo{year}{2020}).

\bibitem[{\citenamefont{Walschaers et~al.}(2020)\citenamefont{Walschaers,
  Parigi, and Treps}}]{trepstheorem}
\bibinfo{author}{\bibfnamefont{M.}~\bibnamefont{Walschaers}},
  \bibinfo{author}{\bibfnamefont{V.}~\bibnamefont{Parigi}}, \bibnamefont{and}
  \bibinfo{author}{\bibfnamefont{N.}~\bibnamefont{Treps}}, ``Practical
  framework for conditional non-{G}aussian quantum state preparation,''
  \bibinfo{journal}{PRX Quantum} \textbf{\bibinfo{volume}{1}},
  \bibinfo{pages}{020305} (\bibinfo{year}{2020}).

\bibitem[{\citenamefont{Brecht and Silberhorn}(2013))}]{brecht}
\bibinfo{author}{\bibfnamefont{B.}~\bibnamefont{Brecht}} \bibnamefont{and}
  \bibinfo{author}{\bibfnamefont{C.}~\bibnamefont{Silberhorn}},
  ``Characterizing entanglement in pulsed parametric down-conversion using
  chronocyclic wigner functions,'' \bibinfo{journal}{Phys. Rev. A}
  \textbf{\bibinfo{volume}{87}}, \bibinfo{pages}{053810}
  (\bibinfo{year}{2013)}).

\bibitem[{\citenamefont{Triginer et~al.}(2020))\citenamefont{Triginer,
  Vidrighin, Quesada, Eckstein, Moore, Kolthammer, Sipe, and
  Walmsley}}]{triginer}
\bibinfo{author}{\bibfnamefont{G.}~\bibnamefont{Triginer}},
  \bibinfo{author}{\bibfnamefont{M.~D.} \bibnamefont{Vidrighin}},
  \bibinfo{author}{\bibfnamefont{N.}~\bibnamefont{Quesada}},
  \bibinfo{author}{\bibfnamefont{A.}~\bibnamefont{Eckstein}},
  \bibinfo{author}{\bibfnamefont{M.}~\bibnamefont{Moore}},
  \bibinfo{author}{\bibfnamefont{W.~S.} \bibnamefont{Kolthammer}},
  \bibinfo{author}{\bibfnamefont{J.~E.} \bibnamefont{Sipe}}, \bibnamefont{and}
  \bibinfo{author}{\bibfnamefont{I.~A.} \bibnamefont{Walmsley}},
  ``Understanding high-gain twin-beam sources using cascaded stimulated
  emission,'' \bibinfo{journal}{Phys. Rev. X} \textbf{\bibinfo{volume}{10}},
  \bibinfo{pages}{031063} (\bibinfo{year}{2020)}).

\bibitem[{\citenamefont{Braunstein and Van~Loock}(2005)}]{contvar1}
\bibinfo{author}{\bibfnamefont{S.~L.} \bibnamefont{Braunstein}}
  \bibnamefont{and}
  \bibinfo{author}{\bibfnamefont{P.}~\bibnamefont{Van~Loock}}, ``Quantum
  information with continuous variables,'' \bibinfo{journal}{Rev. Mod. Phys.}
  \textbf{\bibinfo{volume}{77}}, \bibinfo{pages}{513} (\bibinfo{year}{2005}).

\bibitem[{\citenamefont{Weedbrook et~al.}(2012)\citenamefont{Weedbrook,
  Pirandola, Garc{\'\i}a-Patr{\'o}n, Cerf, Ralph, Shapiro, and
  Lloyd}}]{weedbrook}
\bibinfo{author}{\bibfnamefont{C.}~\bibnamefont{Weedbrook}},
  \bibinfo{author}{\bibfnamefont{S.}~\bibnamefont{Pirandola}},
  \bibinfo{author}{\bibfnamefont{R.}~\bibnamefont{Garc{\'\i}a-Patr{\'o}n}},
  \bibinfo{author}{\bibfnamefont{N.~J.} \bibnamefont{Cerf}},
  \bibinfo{author}{\bibfnamefont{T.~C.} \bibnamefont{Ralph}},
  \bibinfo{author}{\bibfnamefont{J.~H.} \bibnamefont{Shapiro}},
  \bibnamefont{and} \bibinfo{author}{\bibfnamefont{S.}~\bibnamefont{Lloyd}},
  ``Gaussian quantum information,'' \bibinfo{journal}{Rev. Mod. Phys.}
  \textbf{\bibinfo{volume}{84}}, \bibinfo{pages}{621} (\bibinfo{year}{2012}).

\bibitem[{\citenamefont{Adesso et~al.}(2014)\citenamefont{Adesso, Ragy, and
  Lee}}]{contvar2}
\bibinfo{author}{\bibfnamefont{G.}~\bibnamefont{Adesso}},
  \bibinfo{author}{\bibfnamefont{S.}~\bibnamefont{Ragy}}, \bibnamefont{and}
  \bibinfo{author}{\bibfnamefont{A.~R.} \bibnamefont{Lee}}, ``Continuous
  variable quantum information: Gaussian states and beyond,''
  \bibinfo{journal}{Open Syst. Inf. Dyn.} \textbf{\bibinfo{volume}{21}},
  \bibinfo{pages}{1440001} (\bibinfo{year}{2014}).

\bibitem[{\citenamefont{Bloch and Messiah}(1962)}]{blochmessiah}
\bibinfo{author}{\bibfnamefont{C.}~\bibnamefont{Bloch}} \bibnamefont{and}
  \bibinfo{author}{\bibfnamefont{A.}~\bibnamefont{Messiah}}, ``The canonical
  form of an antisymmetric tensor and its application to the theory of
  superconductivity,'' \bibinfo{journal}{Nucl. Phys.}
  \textbf{\bibinfo{volume}{39}}, \bibinfo{pages}{95} (\bibinfo{year}{1962}).

\bibitem[{\citenamefont{Horoshko et~al.}(2019)\citenamefont{Horoshko, La~Volpe,
  Arzani, Treps, Fabre, and Kolobov}}]{horoshko}
\bibinfo{author}{\bibfnamefont{D.~B.} \bibnamefont{Horoshko}},
  \bibinfo{author}{\bibfnamefont{L.}~\bibnamefont{La~Volpe}},
  \bibinfo{author}{\bibfnamefont{F.}~\bibnamefont{Arzani}},
  \bibinfo{author}{\bibfnamefont{N.}~\bibnamefont{Treps}},
  \bibinfo{author}{\bibfnamefont{C.}~\bibnamefont{Fabre}}, \bibnamefont{and}
  \bibinfo{author}{\bibfnamefont{M.~I.} \bibnamefont{Kolobov}},
  ``Bloch-{M}essiah reduction for twin beams of light,''
  \bibinfo{journal}{Phys. Rev. A} \textbf{\bibinfo{volume}{100}},
  \bibinfo{pages}{013837} (\bibinfo{year}{2019}).

\bibitem[{\citenamefont{Roux}(2018)}]{stquad}
\bibinfo{author}{\bibfnamefont{F.~S.} \bibnamefont{Roux}}, ``Combining
  spatiotemporal and particle-number degrees of freedom,''
  \bibinfo{journal}{Phys. Rev. A} \textbf{\bibinfo{volume}{98}},
  \bibinfo{pages}{043841} (\bibinfo{year}{2018}).

\bibitem[{\citenamefont{Roux}(2020{\natexlab{b}})}]{stquaderr}
\bibinfo{author}{\bibfnamefont{F.~S.} \bibnamefont{Roux}}, ``Erratum:
  {C}ombining spatiotemporal and particle-number degrees of freedom [{P}hys.
  {R}ev. {A} 98, 043841 (2018)],'' \bibinfo{journal}{Phys. Rev. A}
  \textbf{\bibinfo{volume}{101}}, \bibinfo{pages}{019903(E)}
  (\bibinfo{year}{2020}{\natexlab{b}}).

\bibitem[{\citenamefont{Mrowczynski and Mueller}(1994)}]{mrowc}
\bibinfo{author}{\bibfnamefont{S.}~\bibnamefont{Mrowczynski}} \bibnamefont{and}
  \bibinfo{author}{\bibfnamefont{B.}~\bibnamefont{Mueller}}, ``Wigner
  functional approach to quantum field dynamics,'' \bibinfo{journal}{Phys. Rev.
  D} \textbf{\bibinfo{volume}{50}}, \bibinfo{pages}{7542}
  (\bibinfo{year}{1994}).

\bibitem[{\citenamefont{Groenewold}(1946)}]{groenewold}
\bibinfo{author}{\bibfnamefont{H.~J.} \bibnamefont{Groenewold}}, ``On the
  principles of elementary quantum mechanics,'' \bibinfo{journal}{Physica}
  \textbf{\bibinfo{volume}{12}}, \bibinfo{pages}{405} (\bibinfo{year}{1946}).

\bibitem[{\citenamefont{Moyal}(1949)}]{moyal}
\bibinfo{author}{\bibfnamefont{J.~E.} \bibnamefont{Moyal}}, ``Quantum mechanics
  as a statistical theory,'' \bibinfo{journal}{Math. Proc. Camb. Philos. Soc.}
  \textbf{\bibinfo{volume}{45}}, \bibinfo{pages}{99} (\bibinfo{year}{1949}).

\bibitem[{\citenamefont{Curtright and Zachos}(2012)}]{psqm}
\bibinfo{author}{\bibfnamefont{T.~L.} \bibnamefont{Curtright}}
  \bibnamefont{and} \bibinfo{author}{\bibfnamefont{C.~K.}
  \bibnamefont{Zachos}}, ``Quantum mechanics in phase space,''
  \bibinfo{journal}{Asia Pacific Physics Newsletter}
  \textbf{\bibinfo{volume}{1}}, \bibinfo{pages}{37} (\bibinfo{year}{2012}).

\bibitem[{\citenamefont{Roux}(2020{\natexlab{c}})}]{nosemi}
\bibinfo{author}{\bibfnamefont{F.~S.} \bibnamefont{Roux}}, ``Parametric
  down-conversion beyond the semiclassical approximation,''
  \bibinfo{journal}{Phys. Rev. Research} \textbf{\bibinfo{volume}{2}},
  \bibinfo{pages}{033398} (\bibinfo{year}{2020}{\natexlab{c}}).

\bibitem[{Note1()}]{Note1}
\bibinfo{note}{Both $A$ and $A^{-1}$ have the identity $\protect \mathbf
  {1}$ as their leading terms. When we replace $A$ and $A^{-1}$ by $\protect
  \mathbf {1}$ in the last two terms in Eq.~(\ref {tmpn}), they cancel.}

\end{thebibliography}
\end{document}